\documentclass{aipproc}

\layoutstyle{6x9}

\SetInternalRegister\hbadness{8000} % pseudo latin isn't breaking very well :-)

\newcommand\doingARLO[2][]{%
  \ifx\mmref\undefined #1\else #2\fi
}

\newfam\msbfam
\batchmode\font\twelvemsb=msbm10 scaled\magstep1 \errorstopmode
\ifx\twelvemsb\nullfont\def\Bbb{\bf}
        \font\fourteenbbb=cmb10 at 14pt
	
	\font\tenbbb=cmb10
\else   \catcode`\@=11
        \font\tenmsb=msbm10 \font\eightmsb=msbm8 
	 
        \textfont\msbfam=\twelvemsb
        \scriptfont\msbfam=\tenmsb \scriptscriptfont\msbfam=\eightmsb
        \def\Bbb{\relax\expandafter\Bbb@}
        \def\Bbb@#1{{\Bbb@@{#1}}}
        \def\Bbb@@#1{\fam\msbfam\relax#1}
        \catcode`\@=\active
	\font\fourteenbbb=msbm10 at 14pt
	
	\font\tenbbb=msbm10
\fi
\newfam\scrfam
\batchmode\font\tenscr=rsfs10 \errorstopmode
\ifx\tenscr\nullfont
        \def\scr{\cal}
\else   \font\eightscr=rsfs10 at 8pt
        \font\sevenscr=rsfs7
        \font\fivescr=rsfs5
        \font\twelvescr=rsfs10 at 12pt
        \skewchar\tenscr='177 \skewchar\sevenscr='177 \skewchar\fivescr='177
        \textfont\scrfam=\twelvescr \scriptfont\scrfam=\tenscr
        \scriptscriptfont\scrfam=\eightscr
        \def\scr{\fam\scrfam}
        \def\cal{\scr}
\fi
\def\hepth#1{\href{http://xxx.lanl.gov/abs/hep-th/#1}{{\xtt hep-th/#1}}}
\def\jhep#1#2#3#4{\href{http://jhep.sissa.it/stdsearch?paper=#2\%28#3\%29#4}{J. High Energy Phys. {\xbold #1#2} ({\xold#3}) {\xold#4}}}
\def\AP#1#2#3{Ann. Phys. {\xbold#1} ({\xold#2}) {\xold#3}}

\def\CQG#1#2#3{Class. Quantum Grav. {\xbold#1} ({\xold#2}) {\xold#3}}

\def\JHEP{\jhep}

\def\NPB#1#2#3{Nucl. Phys. {\xbf B}{\xbold#1} ({\xold#2}) {\xold#3}}

\def\PLB#1#2#3{Phys. Lett. {\xbf B}{\xbold#1} ({\xold#2}) {\xold#3}}
\def\PR#1#2#3{Phys. Rept. {\xbold#1} ({\xold#2}) {\xold#3}}

\def\PRL#1#2#3{Phys. Rev. Lett. {\xbold#1} ({\xold#2}) {\xold#3}}

\def\href#1#2{{#2}}

\def\R{{\Bbb R}}  
\def\Z{{\Bbb Z}}  
\def\bigZ{{\fourteenbbb Z}}

\def\a{\alpha}
\def\b{\beta}
\def\g{\gamma}
\def\d{\delta}
\def\e{\epsilon}
\def\k{\kappa}
\def\l{\lambda}
\def\th{\theta}
\def\w{\omega}
\def\x{\chi}

\def\D{\Delta}
\def\G{\Gamma}

\def\A{{\cal A}}  
\def\C{{\cal C}}  
\def\F{{\cal F}}  
\def\H{{\cal H}}  
\def\L{{\cal L}}
\def\U{{\cal U}} 

\def\tJ{\tilde J} 

\def\Dslash{D\hskip-6.5pt/\hskip1.5pt}

\def\ra{\rightarrow}
\def\lra{\longrightarrow}
\def\arrowunder#1{\raise4pt\vtop{\baselineskip=0pt\lineskip=0pt
      \ialign{\hfill##\hfill\cr${\ss #1}$\cr$\lra$\cr}}}
\def\Darrow#1{\;\arrowunder{\D_{#1}}\;}

\def\small#1{{\hbox{$#1$}}}
\def\*{\partial}
\def\half{\small{1\over2}}
\def\fraction#1{\small{1\over#1}}
\def\fr{\fraction}
\def\Fraction#1#2{\small{#1\over#2}}
\def\Fr{\Fraction}
\def\id{1\hskip-4pt 1}

\def\tr{\hbox{tr}\,}

\def\eg{{\it e.g.}}

\def\ie{{\it i.e.}}

\def\ss{}
\def\xit{}
\def\xbf{\bf}
\def\xbold{\bf}
\def\xold{}
\def\xtt{}
\def\xrm{}
\def\nlni{\newline}
\def\nl{}

\def\be{\begin{equation}}
\def\ee{\end{equation}}
\def\bea{\begin{eqnarray}}
\def\eea{\end{eqnarray}}

\def\punkt{\,\,.}
\def\komma{\,\,,}

\def\II{\hbox{I\hskip-1.5pt I}}

\begin{document}

\title{Superspace Methods in String Theory, Supergravity and Gauge Theory}

\classification{43.35.Ei, 78.60.Mq}
\keywords{Document processing, Class file writing, \LaTeXe{}}

\author{Martin Cederwall}{
  address={Institute for Theoretical Physics,
	G\"oteborg University and 
	Chalmers University of Technology,
	SE-412 96 G\"oteborg, Sweden},
  email={martin.cederwall@fy.chalmers.se},
}

% \copyrightholder{Acoustical Scociety of America}
\copyrightyear{2001}

\begin{abstract}
In these two lectures, delivered at the XXXVII Karpacz Winter School,
February 2001, I review some applications of superspace in
various topics related to string theory and M-theory.
The first lecture is mainly devoted to descriptions of brane dynamics
formulated in supergravity backgrounds.
The second lecture concerns the use of superspace techniques for
determining consistent interactions in supersymmetric gauge theory
and supergravity, {\it e.g.} $\a'$-corrections from string/M-theory.
\end{abstract}

\date{\today}

\maketitle

\section{I. Branes, Supergravity and superspace}

\subsection{I.1 $p$-branes: $D=11$, $p=2$}

Branes play important r\^oles in string theory and M-theory.
They are non-perturbative objects that may be described as
solitons of the low-energy effective supergravity theories (see refs.
\cite{cede-Stelle} and \cite{cede-Duff} for extensive reviews). 
Here, I will concentrate on
the dynamics of branes, as described by their actions 
\cite{cede-Achucarro,cede-BSTMii,cede-DuffLu,cede-HoweSezginSuperBranes,cede-Dp}.
There are a number of different branes in string theory and M-theory,
most conveniently characterised by their field content when seen as
a field theory on the world-volume. The simplest ones, the so-called
$p$-branes, have a scalar multiplet on the world-volume.
D-branes contain a vector multiplet, coupling to string endpoints
\cite{cede-Polchinski},
and the M5-brane has a self-dual tensor.

As a model for the simplest branes I will treat the membrane in 
eleven dimensions \cite{cede-BSTMii}. The action for a brane typically
consists of two parts, a kinetic term proportional to the 
invariant volume, and a Wess--Zumino term specifying the minimal
coupling to a $(p+1)$-form potential, under which the brane carries
charge:
\be
S=-T\int d^3\xi\sqrt{-|g|}+T\int C\punkt
\label{cede-MembraneAction}
\ee
$T$ is the membrane tension.
The action (\ref{cede-MembraneAction}) looks like an action describing bosonic
degrees of freedom contained in its transverse fluctuations.
How do we describe a supersymmetric brane, containing an equal number
of fermionic degrees of freedom? One simple and very efficient way,
in many aspects much simpler than a component approach, is to
consider the dynamics to be described by the same formal action,
but where the bosonic world-volume is embedded in 
superspace\footnote{There exists a framework, the so called embedding
formalism, where both target space and the world-volume are superspaces
\cite{cede-HoweSezginSuperBranes}.
I will not consider it here.}. The target superspace has coordinates
$Z^M=(X^m,\th^\mu)$ (the corresponding inertial indices are $A=(a,\a)$,
a Lorentz vector and some spinor), and the background fields entering
the action (\ref{cede-MembraneAction}) are pullbacks from superspace to
the world-volume, 
$g_{ij}=E_i{}^aE_j{}^b\eta_{ab}$, 
$C_{ijk}=E_k{}^CE_j{}^BE_i{}^aC_{ABC}$,
%\bea
%g_{ij}&=&E_i{}^aE_j{}^b\eta_{ab}\komma\\
%C_{ijk}&=&E_k{}^CE_j{}^BE_i{}^aC_{ABC}\komma
%\eea
with $E_i{}^A=\*_iZ^ME_M{}^A$, $E_M{}^A$ being the target space
super-vielbein.

Let us now investigate what this means for the supermembrane. The
eight transverse bosonic oscillations must be matched in number by
eight fermionic degrees of freedom if the action is to be supersymmetric.
A (Majorana) spinor in $D=11$ has 32 components. The number is reduced
by half by the equations of motion, as usual, but it is clear that
an additional local symmetry is required in order to get eight physical
spinor degrees of freedom. This is the so called $\k$-symmetry, parametrised
by a half spinor. $\k$-symmetry is a local (in terms of the location on
the brane) translation of the brane in a fermionic direction in
superspace. As such, it is generated by a superspace vector field pointing
in fermionic directions only: $\k=\k^M\*_M=\k^\a E_\a{}^M\*_M$,
and the transformation of the coordinates is
$\d_\k Z^M=\k^M$. Pullbacks of superspace forms are transformed by
the Lie derivative,
$\d_\k f^*\Omega=f^*\L_\k\Omega=f^*(i_\k d+di_\k)\Omega$,
%\be
%\d_\k f^*\Omega=f^*\L_\k\Omega=f^*(i_\k d+di_\k)\Omega\komma
%\ee
which after a brief calculation implies that (pullbacks are suppressed
in the following)
\bea
\d_\k C&=&i_\k H+di_\k C\komma\\
\d_\k E^A&=&D\k^A+i_\k T^A\komma\\
\d_\k g_{ij}&=&2E_{(i}{}^aE_{j)}^B\k^\a T_{\a B}{}^b\eta_{ab}\komma
\eea
where $H=dC$ is the background tensor superfield strength and $T^A=DE^A$
the superspace torsion.

To determine how the action transforms (modulo boundary terms), we only need
$i_\k H$ and $i_\k T^a$. In $D=11$ supergravity 
\cite{cede-ElevenSG,cede-ElevenSSSG} this is particularly simple,
the only non-vanishing components of $H$ and $T^a$ with at least one
spinorial form-index are the dimension 0 ones,
$T_{\a\b}{}^a=2\G_{\a\b}^a$,
$H_{ab\a\b}=2(\G_{ab})_{\a\b}$.
%\bea
%T_{\a\b}{}^a&=&2\G_{\a\b}^a\komma\label{cede-TorsionConstraint}\\
%H_{ab\a\b}&=&2(\G_{ab})_{\a\b}\punkt
%\eea
A short calculation yields
%\bea
%\d_\k S&=&-\int d^3\xi\half\sqrt{-g}g^{ij}\d_\k g_{ij}+\int\d_\k C\nonumber\\
%&=&-\int d^3\xi 2\sqrt{-g}
%	E_i{}^\a(\G^i-\fr{2\sqrt{-g}}\e^{ijk}\G_{jk})_{\a\b}\k^\b\punkt
%\eea
\be
\d_\k S=-\int d^3\xi 2\sqrt{-g}
	E_i{}^\a(\G^i-\fr{2\sqrt{-g}}\e^{ijk}\G_{jk})_{\a\b}\k^\b\punkt
\ee
with the obvious notation for pullbacks of $\G$-matrices.
The combination of $\G$-matrices in the last term may be written as
\be
\G^i-\fr{2\sqrt{-g}}\e^{ijk}\G_{jk}=\G^i(\id-\fr{6\sqrt{-g}}\e^{ijk}\G_{ijk})
=\G^i(\id-\G)\komma\label{cede-MembraneGamma}
\ee
and is seen to provide a projection on $\k$, since
$\Pi_\pm=\half(\id\pm\G)$, due to the identities $\G^2=\id$ and $\tr\G=0$, are
projection matrices splitting a 32-component spinor in two halves.
The only chance that this variation vanishes is thus that
$\Pi_-\k=0$. This is indeed the half spinor of local fermionic symmetry
that was needed for the matching of bosonic and fermionic degrees of freedom.
Since setting the dimension 0 torsion to a $\G$-matrix puts the background
on shell \cite{cede-HoweWeyl}, the supermembrane has $\k$-symmetry in any
on-shell background of $D=11$ supergravity.

Analogous calculations hold for other $p$-branes in other supergravities,
and show that for general on-shell backgrounds, the actions are
$\k$-symmetric. $\k$-symmetry is related to the fact that the branes
are BPS-saturated configurations---the supersymmetry algebra generating
the multiplets on the branes (in the present case a scalar multiplet)
contains half the number of fermionic generators compared to the target space
supersymmetry, and half of the target space supersymmetry is broken
(the world-volume fields are Goldstone fields corresponding to broken
symmetries of the background). The projection matrices are related to
(target space) supersymmetry algebras with ``central'' tensorial charges,
that get projected out by a half-rank projection $\Pi_\pm$.

We may also note that the formalism presented here, with the brane
embedded in an arbitrary target superspace background, actually is
as simple as in a flat superspace. Working with explicit fermionic
coordinates becomes complicated, since the expression for a tensor
potential is complicated, while the (gauge invariant) field strength
is simple.

As presented here, the branes are viewed as infinitely thin objects
moving in superspace. They may also be seen as solitons in the
low-energy effective supergravity theories \cite{cede-DuffKhuriLu}.
All fields on branes arise as Goldstone modes corresponding to
broken symmetries of the background theory. Scalars and fermions
correspond to broken translational symmetries and supersymmetry,
while vectors on D-branes and tensors on M5-branes arise as Goldstone
modes for large gauge symmetries of target space tensors, \ie, gauge
transformations that take different values ``on the brane'' and in
the asymptotic region \cite{cede-ACGNR}.

\subsection{I.2 $D=11$ supergravity}

The $\G$-matrix constraint on the dimension 0 torsion
puts the theory on shell \cite{cede-HoweWeyl}. The tensor field arises
naturally from the superspace geometry, and it is not necessary to
separately require the existence of a closed 4-form on superspace.
The Bianchi identity for $H$ at dimension 0 becomes
$0=(dH)_{a\a\b\g\d}=6T_{(\a\b}{}^bH_{|ba|\g\d)}
	=24\G_{(\a\b}^b\G^{\phantom{b}}_{|ba|\g\d)}$,
%\be
%0=(dH)_{a\a\b\g\d}=6T_{(\a\b}{}^bH_{|ba|\g\d)}
%	=24\G_{(\a\b}^b\G^{\phantom{b}}_{|ba|\g\d)}\komma
%\ee
which is fulfilled due to a Fierz identity in eleven dimensions,
and at dimension 1, the non-vanishing torsion is
\be
T_{a\a}{}^\b=\fr{36}\G^{bcd}{}_\a{}^\b H_{abcd}
	+\fr{288}\G_a{}^{bcde}{}_\a{}^\b H_{bcde}\punkt
\ee
Actually, the superspace Bianchi identities also leave room for
a spinor $\omega_\a$ at dimension $\half$ and a vector $\omega_a$
at dimension 1 in $T$, the Bianchi identities further require that
these be integrable to $\omega_A=D_A\phi$, and the ``conformal compensator''
$\phi$ can then be removed by a conventional constraint, or alternatively
by the enlargement of the structure group to include 
Weyl rescalings\footnote{There is a disagreement on this point. The
view presented here is that of refs. \cite{cede-HoweWeyl,cede-CGNN}, while the
authors of ref. \cite{cede-GatesNishino} claim that the conformal compensator
has to play a r\^ole in a (yet unknown) supersymmetric 
off-shell formulation of
eleven-dimensional supergravity.}.

\subsection{I.3 D-branes, type \II\ supergravity}

Type \II\ superstring theories, and their low-energy effective theories,
type \II A and \II B supergravity, contain tensor fields in the
Ramond-Ramond sector. For type \II A the potentials have odd rank, 
$C=C_{(1)}\oplus C_{(3)}\oplus C_{(5)}\oplus\ldots$, and for
type \II B even, $C=C_{(0)}\oplus C_{(2)}\oplus C_{(4)}\oplus\ldots$.
The corresponding field strengths are required to be self-dual, so in
principle we have a redundant set of potentials, which is useful when
considering brane actions. A five-brane, \eg, couples minimally
to a 6-form potential, whose 7-form field strength is dual to
the 3-form. In addition there is the the NS-NS 2-form $B$.

D-branes are exactly the non-perturbative objects carrying charge
under the RR fields. They act as hypersurfaces where fundamental strings
are allowed to end, and contain vector degrees of freedom, coupling
minimally to the world-lines of the string ends \cite{cede-Polchinski}.
This picture resulted in an effective action for D-branes
\cite{cede-Leigh,cede-Douglas}:
\bea
S&=&-\int d^{p+1}\xi e^{-\phi}\sqrt{-|g_s+F|}+\int e^FC\nonumber\\
&=&-\int d^{p+1}\xi e^{{p-3\over4}\phi}\sqrt{-|g_E+e^{-\phi/2}F|}
	+\int e^FC\punkt\label{cede-DbraneAction}
\eea
There are some things to explain in this expression. The field $\phi$
is the dilaton field, and the factor $\exp(-\phi)$ means that the D-brane
tension in the ``string frame'' is proportional to $g^{-1}$, 
where $g=\exp(\phi)$ is the string coupling. 
In the second line, the action has been rewritten in terms of the 
Einstein metric $g_E=\exp(\phi/2)g_s$, which is sometimes convenient,
especially when I later want to consider SL(2;$\Z$) duality symmetry.
The second, Wess--Zumino, term in the action is evaluated with wedge
products, and the $(p+1)$-form is extracted, so that for the D3-brane, \eg,
it reads $\int(C_{(4)}+F\wedge C_{(2)}+\half F\wedge F\wedge C_{(0)})$.
There is a U(1) vector field $A$ on the world-volume, and the field strength
$F$ contains the NS-NS 2-form potential $B$ trough $F=dA-B$. The
gauge transformations of $B$, $\d_{\l}B=d\l$, also act on $A$ as
$\d_{\l}A=\l$, so that $F$ is invariant. This means that an expectation
value for $F$ can be traded for a 
background $B$ field\footnote{This implies that such configurations
do not break supersymmetry, which can be seen in the supergravity
solutions corresponding to D-branes with constant $F$, or equivalently,
in a background $B$ field \cite{cede-FiniteFSols}. 
What instead happens is that the part of
supersymmetry remaining unbroken is a different projection than for $F=0$,
due to the $F$-dependence of the projector (\ref{cede-DbraneGamma}).}. 
Apart from the
dilaton factor, the first, kinetic, term is of Dirac--Born--Infeld type.

The RR tensors have ``modified'' field strengths
$R=e^Bd(e^{-B}C)=dC-H\wedge C$, and their Bianchi identities read
$dR+H\wedge R=0$. Gauge transformations $\d_\Lambda C=e^Bd\Lambda$
leave the WZ term invariant up to a total derivative.

As was done for the membrane in the previous section, the D-brane
actions (\ref{cede-DbraneAction}) are promoted to actions for supersymmetric
D-branes by letting the the embedding be in a target superspace of
type \II A or \II B. Superspace formulations of the type \II A and
type \II B supergravities are given in refs. \cite{cede-CarrGatesOerterIIA}
and \cite{cede-HoweWestIIB}.

The essential check is again $\k$-symmetry. The calculations are somewhat
more complicated than for a brane with a scalar multiplet, so I refer to
ref. \cite{cede-Dp} for more details. In type \II B, the two spinor coordinates
have the same chirality, and instead of introducing explicit indices,
I include this in a \II B spinor index $\a=1,\ldots,32$, and introduce
the basis of $2\times2$-matrices $\{\id,I,J,K\}$  with 
$I=i\sigma_2$, $J=\sigma_1$, $K=\sigma_3$ (they can be seen as a basis 
for the split quaternions).
The the relevant fields at dimension 0 are
\bea
T_{\a\b}{}^a&=&2\G_{\a\b}^a\komma\\
H_{a\a\b}&=&-2e^{\phi\over2}(\G_aK)_{\a\b}\komma\\
R_{a_1\ldots a_{n-2}\a\b}
	&=&2e^{{n-5\over4}\phi}(\G_{a_1\ldots a_{n-2}}K^{n-1\over2}I)_{\a\b}
\punkt
\eea
Since the ten-dimensional supergravities have spinors, $\l_\a=D_\a\phi$
of dimension $\half$, these will also occur in the dimensions $\half$
components of $H$, $R$ and $T$, which I do not list here.

The variation of the action requires that we specify the transformation
of the vector $A$. In order to get something gauge invariant we must take
$\d_\k A=i_\k B$, which implies that $\d_\k F=-i_\k H$.
Then the variation of the WZ term becomes
$\d_\k(e^FC)=e^Fi_\k R$ modulo boundary terms.
Going through the procedure of inserting the variations of the fields in
the lagrangian yields an expression 
\be
\d_\k L\propto E_i{}^\a\G^i(\id-\G)\k^\a\komma
\ee
where $\G$ is a more complicated expression than eq. (\ref{cede-MembraneGamma}),
containing different powers of the field strength $F$, and thus providing
field-dependent half-rank projections of a spinor.
For the D3-brane, it takes the form
\be
\G={\e^{ijkl}\over\sqrt{-|g+e^{-\phi/2}F|}}
	\left(\fr{24}\G_{ijkl}I-\fr4e^{-{\phi\over2}}F_{ij}\G_{kl}J
	+\fr8e^{-\phi}F_{ij}F_{kl}I\right)\komma\label{cede-DbraneGamma}
\ee
and similar expressions hold for other D-branes.
This shows $\k$-symmetry of the D-branes in an arbitrary on-shell
supergravity background.

\subsection{I.4 SL(2;\bigZ), tensor democracy}

Type \II B supergravity has an SL(2;$\R$) symmetry, which at the quantum
level is broken to the SL(2;$\Z$) S-duality group. Since S-duality is
non-perturbative, the representations under SL(2;$\Z$) contain 
perturbative and non-perturbative states, and can not be manifested
in perturbative string theory. Nevertheless, it can be manifested
in the effective supergravity, and it is meaningful to ask whether
it is possible to treat all branes with NS-NS and RR charges, including
the fundamental string, on an equal footing, thus manifesting 
the S-duality symmetry.

The scalars, the dilaton and axion, belong to the coset SL(2;$\R$)/U(1).
This is a combination of NS-NS and RR fields, since the axion is
identified with $C_{(0)}$. The NS-NS and RR 2-forms $B_{(2)}$ and
$C_{(2)}$ combine into an SL(2) doublet, and the 4-form (with
selfdual field strength) is an SL(2) singlet. Higher rank tensors
are dual to those already mentioned.
The representations of branes reflect those of the tensor fields:
The strings and five-branes 
come with charges that form an SL(2;$\Z$) doublet $(p,q)$,
while the D3-brane forms a singlet. I will not examine higher-dimensional
branes here.

The scalars of type \II B supergravity are describe as a complex 
doublet $\U^r$, $r=1,2$, subject to the constraint 
$\Fr i2\e_{rs}\U^r\bar\U^s=1$ (which is the condition that $\U$ has unit 
determinant when seen as a real $2\times2$ matrix).
The coset is obtained from gauging the U(1) acting as 
$\U\rightarrow e^{i\th}\U$. One forms the left-invariant Maurer--Cartan
forms
%$Q=\half\e_{rs}d\U^r\bar\U^s$,
%$P=\half\e_{rs}d\U^r\U^s$,
\bea
Q&=&\half\e_{rs}d\U^r\bar\U^s\komma\\
P&=&\half\e_{rs}d\U^r\U^s\komma
\eea
with Bianchi identities (Maurer--Cartan equations)
%$dQ-iP\wedge\bar P=0$,
%$DP\equiv dP-2iP\wedge Q=0$.
\bea
dQ-iP\wedge\bar P&=&0\komma\\
DP\equiv dP-2iP\wedge Q&=&0\punkt
\eea
The scalars act as a bridge between objects that are SL(2) doublets and
real and objects that are SL(2) singlets but carry U(1) charge.
If we write the 3-form doublet of field strengths as
$H_{(3)r}=dC_{(2)r}$, the SL(2) singlet field strength is
$\H_{(3)}=\U^rH_{(3)r}$. Notice that this is necessary when writing
a kinetic term as proportional to $\H\cdot\bar\H$.
The Bianchi identity is $D\H+i\bar\H\wedge P=0$ (recall that $\H$ has U(1)
charge 1 while P has charge 2).
The singlet 5-form is constructed as
$H_{(5)}=dC_{(4)}+\hbox{Im}(\C_{(2)}\wedge\bar\H_{(3)})$, with Bianchi
identity $dH_{(5)}-i\H_{(3)}\wedge\bar\H_{(3)}=0$.

We now come to the crucial point in describing brane dynamics 
SL(2)-covariantly. It is not sufficient to introduce one vector field
on the brane. Remember that the field strength was $F=dA-B$, where 
$B$ was the NS-NS 2-form. It is clear that another vector, combining
with the RR 2-form is needed, so that they form a doublet. One
should thus have $F_r=dA_r-C_{(2)r}$, reflecting the fact that strings
of different charges $(p,q)$ can end on a brane.
Once this step has been taken, it is equally natural to introduce
a form of rank $p$ for each  background tensor fields
of rank $p+1$, reflecting the fact
that a $p$-brane can end on the brane we describe, and coupling
minimally to its boundary. For this reason, such a formulation,
with complete ``tensor democracy'' on the branes, should
most naturally encode the coupling of branes to background fields.
Gauge invariance (in target space and on the brane) demands that
also the tensors on the brane have modified Bianchi identities.
The 2-form and 4-form on any brane are 
\bea
\F_{(2)}&=&\U^rdA_{(1)r}-\C_{(2)}\komma\\
F_{(4)}&=&dA_{(3)}-C_{(4)}+\hbox{Im}(\A_{(1)}\wedge\bar\H_{(3)})\komma
\eea
with Bianchi identities
\bea
D\F_{(2)}+i\bar\F_{(2)}\wedge P&=&-\H_{(3)}\komma\\
dF_{(4)}&=&-H_{(5)}-\hbox{Im}(\F_{(2)}\wedge\bar\H_{(3)})\punkt
\eea

In a generic situation, the procedure seems to give too many bosonic fields,
and there must be ways to reduce the number in order to recover
an SL(2)-covariant description of brane dynamics. The key is selfduality,
and I will sketch how it works for different branes. I refer
the readers to refs. \cite{cede-CT,cede-CW} for details.
All cases described may be shown to be $\k$-symmetric, along similar lines
as in the previous sections. The actions do not divide into Born--Infeld
plus Wess--Zumino, since this presumes a division into NS-NS and RR fields.

\paragraph{The $(p,q)$ strings}
The vectors $A_r$ have no local degrees of freedom on the two-dimensional
world-sheet, so we do not have to worry about removing degrees of freedom.
The only degrees of freedom of vectors is a quantised electric flux
(see ref. \cite{cede-WittenDbranes} for one vector), so the description
gives rise to a pair of integers $(p,q)$, which are the charges of string.
In this way, the whole spectrum of $(p,q)$ strings is described
within one single action \cite{cede-CT}. That the description is correct is checked
by $\k$-symmetry and by the fact that the correct tensions 
\cite{cede-SchwarzSL} are produced.

\paragraph{The 3-brane}
Having two vector potentials gives too many degrees of freedom, and
one of them has effectively to be removed. This is obtained by imposing
a selfduality relation on the complex field strength $\F$:
$\F=i\star\F$ + higher order terms. It turns out that not any 
non-linear selfduality relation is allowed. Its exact form is dictated
by consistency with the coupling to the background fields, and also,
independently, by $\k$-symmetry, and it encodes in a manifest way
the earlier observed Poincar\'e selfduality of the 3-brane.  
A formulation of the dynamics of the type \II B 3-brane is obtained \cite{cede-CW}
that naturally encodes in a most symmetric way all couplings to 
background fields, and thereby the possibilities for the 3-brane 
to host brane boundaries \cite{cede-CW}.  

\paragraph{The $(p,q)$ 5-branes}
This case is not constructed in detail, but the general scheme is
described in ref. \cite{cede-CW}. There is a duality relation between
the 4-form and the 2-forms. The fact that the corresponding supergravity
solution could be described analytically \cite{cede-CGNNpq} makes it
reasonable to believe that the dynamics can be described covariantly,
in spite of problems with dualisation in six dimensions
\cite{cede-SchwarzSixDual}.

\paragraph{The M5-brane and type \II A} 
The formalism is not restricted to type \II B.
It was successfully applied to write down a ``quasi-action'' (the equations
of motion follows, but not the selfduality, which however is uniquely
determined by consistency with background couplings and by $\k$-symmetry) 
for the M5-brane \cite{cede-CNS}.
It is also applicable to type \II A branes, and will also there encode
the background interactions in the most natural way.

\subsection{I.5 Summary}

I have described brane dynamics by embedding in superspace,
given a detailed account of the mechanisms behind $\k$-symmetry and
focussed on the couplings of branes to fields in the background
effective supergravity. 

It is known that the effective supergravity theories following from
string theory or M-theory receive corrections to higher order in $\a'$
than the lowest order ones used in this talk. 
Some $\a'$-corrections to the brane actions themselves are also known
\cite{cede-BBG,cede-Wyllard}. What happens to the brane dynamics when 
$\a'$ corrections are turned in in target space? It is clear that
a superspace formulation is desirable in order to answer such questions.
In the following lecture I will describe some recent progress in
the superspace formulations of $D=11$ supergravity and $D=10$ 
super-Yang--Mills theory, both relevant for string/M-theory.

\section{II. String/M-theory corrections to supergravity and super-Yang--Mills}

\subsection{II.1 $D=11$ supergravity cont'd}

We will now continue the discussion of the superspace formulation
of eleven-dimensional supergravity in superspace
\cite{cede-ElevenSSSG,cede-HoweWeyl,cede-CGNN}.
The vielbeins are $E^A=dZ^ME_M{}^A$, and the resulting torsion 2-form is
$T^A=DE^A=dE^A+E^B\wedge\w_B{}^A$, where the structure group is the
Lorentz group, \ie, the spin connection satisfies
$\w_\a{}^\b=\fr4(\G^a{}_b)_\a{}^\b\w_a{}^b$.
The curvature is $R_A{}^B=d\w_A{}^B+\w_A{}^C\wedge\w_C{}^B$.
The Bianchi identities for torsion and curvature are
$DT^A=E^B\wedge R_B{}^A$ and $DR_A{}^B=0$. Of these, one needs only to
use the first one.

As long as torsion and curvature are constructed from vielbeins and
spin connections, the Bianchi identities are automatically fulfilled.
In order to reduce the enormous amount of fields contained in these,
one has however to impose ``conventional constraints'' 
%\cite{cede-ConvConstr} 
connecting
the different components. Then the Bianchi identities become 
integrability conditions that have to be checked, and which imply the
equations of motion (this is true for the maximally supersymmetric
theories I deal with in this lecture).

The conventional constraints are of two types.
The first one uses the freedom in the definition of the torsion
to shift it into the spin connection when possible. These constraints
do not eliminate all of the torsion (as it does in bosonic gravity),
but have the effect of determining the spin connection if terms
of the vielbein, which is desirable.
The second type uses a redefinition of the tangent bundle,
$E^A\rightarrow E^BM_B{}^A$, while keeping the spin connection, and
thus the curvature, invariant (although their components vary due to
the change of basis). We want to use this freedom to the extent that
it enables us to express all vielbein components in terms of the
dimension $-\half$ one, $E_\mu{}^a$. 
%So for example may one use
%$M_a{}^b$ to eliminate $E_a{}^b$ as an independent field to avoid the
%doubling occurring due to $E_\mu{}^a=e_\mu{}^a+(\G^m\th)_\mu e_m{}^a+\ldots$.

Let us now examine the lowest-dimensional torsion components, $T_{\a\b}{}^a$
at dimension 0. I already mentioned that putting it equal to a
$\G$-matrix takes the theory on-shell, so in order to incorporate 
corrections to the ordinary supergravity this constraint (which is not
a conventional constraint) has to be modified.
A general expansion yields, since the torsion is symmetric in the
spinor indices,
\be
T_{\a\b}{}^c=2(\G_{\a\b}^dX_d{}^c
	+\half\G_{\a\b}^{d_1d_2}X_{d_1d_2}{}^c
	+\fr{5!}\G_{\a\b}^{d_1\ldots d_5}X_{d_1\ldots d_5}{}^c)
\ee
Decomposing into irreducible representations of the Lorentz group,
we find that the three ``X-tensors'' contain
$((20000)\oplus(01000)\oplus(00000))
\oplus((11000)\oplus(00100)\oplus(10000))
\oplus((10002)\oplus(00002)\oplus(00010))$.
where standard Dynkin labels for highest weights are used.

If this representation content is compared to the one in the 
dimension-0 matrices for redefining the vielbein, $M_a{}^b$ and $M_\a{}^\b$,
which is $((20000)\oplus(01000)\oplus(00000))
\oplus((00002)\oplus(00010)\oplus(00100)\oplus(01000)\oplus(10000)
\oplus(00000))$, we see that the only remaining components are
$X_{d_1d_2}{}^c|_{(11000)}$ and $X_{d_1\ldots d_5}{}^c|_{(10002)}$,
\ie, the ``irreducible hooks'' \cite{cede-BNLXVI,cede-CGNN}. 
These superfields should encode which
the corrections to the supergravity are, and the equations of motion
for any version of $D=11$ supergravity should follow from the solution 
of the Bianchi identities with a suitable choice of these tensors.
This will be even clearer when we consider spinorial cohomology in a
little while.

Solving the Bianchi identities turns out to be quite complicated,
and we have not succeeded in doing it in full generality.
In ref. \cite{cede-CGNN}, we were able to show that the gravitino equation
of motion received a correction, by solving the Bianchi identities up
to dimension $\Fr32$, encountering on the way some remarkable
numerical coincidences. We found no contribution to the Weyl curvatures
up to this level, which means that the elimination of the conformal
compensator by a conventional constraint is still valid.

\subsection{II.2 $D=10$ super-Yang--Mills}

The study of the general superspace formulation of $D=10$ super-Yang--Mills
is motivated by its connection to string theory and the relevance
for finding non-abelian analogues of the Born--Infeld action
\cite{cede-Tseytlin,cede-OtherTalks}.
An advantage with the system is that it is much easier to analyse
than $D=11$ supergravity, so we hoped that it would be more manageable.

I now work in flat superspace, with $T_{\a\b}{}^a=2\G_{\a\b}^a$ and the
rest of the torsion vanishing. The gauge potential is a superspace
1-form with components $A_A=(A_a,A_\a)$, and the field strength is
$F=dA+A\wedge A$ with Bianchi identity $DF=0$. In components, the
Bianchi identity reads:
\bea
\hbox{dim. }\Fr32:&\qquad&D_{(\a}F_{\b\g)}+2\G_{(\a\b}^cF_{|c|\g)}=0\komma\\
2:&\qquad&2D_{(\a}F_{\b)c}+D_cF_{\a\b}+2\G_{\a\b}^dF_{dc}=0\komma\\
\Fr52:&\qquad&D_\a F_{bc}+2D_{[b}F_{c]\a}=0\komma\\
3:&\qquad&D_{[a}F_{bc]}=0
\eea

Taking $F_{\a\b}=0$ puts the theory on-shell \cite{cede-NilssonSYM}, and
it must be relaxed if we want to incorporate corrections.
The general expansion is 
\be
F_{\a\b}=\G_{\a\b}^aJ_a+\fr{5!}\G_{\a\b}^{a_1\ldots a_5}J_{a_1\ldots a_5}\punkt
\ee
The vector can always be set to zero as a conventional constraint, to
eliminate the ``extra'' vector potential occurring at $\th$-level in
$A_\a$. We then have $A_a=-\fr{32}\G^{\a\b}_aD_\a A_\b$ (in the abelian case).
The relevant deformation lies in the five-form, which is automatically
(anti-)selfdual, due to the chirality of the spinors.
In reference \cite{cede-CederwallNilssonTsimpisI} we were able to solve the
Bianchi identities completely for arbitrary $J$, whose components act
as a super-current multiplet, and obtain the equations of motion,
\bea
0&=&D^bF_{ab}-\l\G_a\l-8D^bK_{ab}+36w_a-\Fr43\{\l,\tJ_a\}
-2\tJ_b\G_a\tJ^b	\nonumber\\
&&\qquad+\fr{140\cdot3!}\tJ_{bcd}\G_a\tJ^{bcd}+\fr{42}[K_{bcde},J_a{}^{bcde}]
	+\fr{42\cdot4!}[D^fJ_{fbcde},J_a{}^{bcde}]\komma
		\label{cede-DeformedEOM}\\
0&=&\Dslash\l-30\psi+\Fr43D^a\tJ_a
	+\Fr5{126\cdot5!}	%-\fr{3024}
	\G^{abcde}[\l,J_{abcde}]\punkt\nonumber
\eea
Apart from $F_{ab}$ and $\l^{\a}$ (which appears in the field strength
as $\l^\a=\fr{10}\G^{a\,\a\b}F_{a\b}$), 
the quantities appearing in these equations all arise, as 
explained in ref. \cite{cede-CederwallNilssonTsimpisI}, 
in the $\theta$ expansion of 
$J_{abcde}$; $\tilde{J}'s$ at first,
$K's$  at second, $\psi$ at third, and $\omega$ at fourth order in $\theta$. 
Explicitly,
their precise relations to $J_{abcde}$ are given by
\bea
\tJ_a&=&\fr{1680}\G^{bcde}DJ_{bcdea}\komma\\
\tJ_{abc}&=&-\fr{12}\G^{de}DJ_{deabc}-\fr{224}\G_{[ab}\G^{defg}DJ_{|defg|c]}
	\komma\\
\tJ_{abcde}&=&DJ_{abcde}+\Fr56\G_{[ab}\G^{fg}DJ_{|fg|cde]}
	+\fr{24}\G_{[abcd}\G^{fghi}DJ_{|fghi|e]}\komma
\eea
\bea
K_{ab}&=&\fr{5376}(D\G^{cde}D)J_{cdeab}\komma\\
K_{abcd}&=&\fr{480}(D\G_{[a}{}^{fg}D)J_{|fg|bcd]}\komma
\eea
\be
\psi_\a=-\fr{840\cdot3!\cdot5!}		%-\fr{604800}
	\G_{abc}{}^{\b\g}\G_{de\,\a}{}^\d D_{[\b}D_\g D_{\d]}
	J^{abcde}\komma
\ee
and finally
\be
w_a=\fr{4032\cdot4!\cdot5!}\G^{[\a\b}_{abc}\G^{\g\d]}_{def}
	D_\a D_\b D_\g D_\d J^{bcdef}\punkt
\ee

I will soon show how one may use this formalism to deduce possible
forms of $\a'$-corrections allowed by supersymmetry. The idea is thus
to take advantage of the fact (normally considered as a drawback)
that the superspace formulation takes the theory on-shell.

\subsection{II.3 Fields and deformations from spinorial cohomology}

Before becoming more specific about string-related corrections to
super-Yang--Mills theory, I would like to digress on an amusing
mathematical structure that has something to tell about maximally
supersymmetric theories.

The basic idea is that the theories we consider are gauge theories, and
that, in a superspace formulation, where all potentials and field strengths
are forms on superspace, all components except the purely spinorial ones
are redundant. Since all physical fields are contained in the objects 
carrying spinorial form indices only, it is interesting to examine the
structure arising from these.
Our complexes are of the form
\be
r_{0}\Darrow0r_{1}\Darrow1r_{2}\Darrow2\ldots\Darrow{n-1}r_{n}\Darrow n\ldots
\komma\label{cede-Complex}
\ee
where $r_{p}$, for some $p\geq0$, 
is the representation carried by a gauge transformation,
$r_{p+1}$ that of a potential and $r_{p+2}$ that of a field strength.
I will refer to the representations $r_{n}$ as $n$-forms, a notation
not to be confused with that of a tensor antisymmetric in vector indices.
The exact definitions are given, both for gauge theory and supergravity,
in the following sections, where it will also be clear why $\D$
is a nilpotent operator. The r\^ole of $r_{p+3}$ is as a Bianchi identity.

Let me describe in more detail how the complexes work, with the
super-Yang--Mills theory as an example. 
We have already seen that $A_\a$ contains the fields of the theory.
The relevant part of the field strength, as argued above,
lies in (00020)\footnote{I use standard Dynkin labels for SO(1,9)}, 
and does not contain $A_a$.
We also note \cite{cede-NilssonSYM,cede-CederwallNilssonTsimpisI}
that part of the dimension-$\Fr32$ Bianchi identity
states the vanishing of the (00030) component of $D_\a F_{\b\g}$.
These observations make it natural to consider, 
not the sequence of completely symmetric
representations in spinor indices, but a restriction of it, namely 
the sequence of Spin(1,9) representations $r_n\equiv(000n0)$.
They are the part of the totally symmetric
product of $n$ chiral spinors that has vanishing $\G$-trace, and may
be represented tensorially as 
$C_{\a_1\ldots\a_n}=C_{(\a_1\ldots\a_n)}$, 
$\G_a{}^{\a_1\a_2}C_{\a_1\a_2\a_3\ldots\a_n}=0$.
For $n=2$, $C$ is an anti-selfdual five-form, for $n=3$ a $\G$-traceless
anti-selfdual five-form spinor, etc.

The operator $\D_n$: $r_n\lra r_{n+1}$ can schematically be written
as $\D_nC_n=\Pi(r_{n+1})DC_n$, where $D$ is the exterior covariant derivative
$D=d\theta^\a D_\a$ and $\Pi(r_n)$ is the algebraic projection
from $\otimes^n_s(00010)$ to $(000n0)$. It is straightforward to write
an explicit tensorial form for $\D$ by subtracting $\G$-traces from $DC$,
but it will not be used here.
It is also straightforward to show that, for an abelian gauge group
and standard flat superspace,
the sequence (\ref{cede-Complex}) forms
a complex, \ie, that $\D^2=0$. This follows simply from the fact that
while $\{D_\a,D_\b\}=-T_{\a\b}{}^cD_c$, the torsion only has a component
$2\G_{\a\b}{}^c$ which is projected out by $\Pi(r_n)$.
This means that for non-abelian gauge theory the complex
should be considered in a flat background, and the deformations
yielded are infinitesimal.

We would now like to calculate the cohomology 
$\H^n=\hbox{Ker}\D_n/\hbox{Im}\D_{n-1}$
of the complex associated
with $D=10$ super-Yang--Mills. This can be done by considering
the decomposition into irreducible representations of the representation
sitting at level $\ell$ in $r_n$, $r_{n}^\ell\equiv\wedge^\ell S\otimes
r_{n}$. This is easily done, \eg\ with the help of the program LiE 
\cite{cede-LiE}.
One then follows each of the irreducible representations at a given
dimension through the subcomplex
$$
r_0^\ell\ra r_1^{\ell-1}
\ra r_2^{\ell-2}
\ra\ldots\ra r_{\ell-1}^{1}\ra
r_\ell\punkt
$$

Let me illustrate the calculation by examining the field content.
We then look into the spinor potential of dimension $\half$, 
which contains all fields
in the vector multiplet, so we should examine the first cohomology.
The vector (dimension 1) sits at
$\ell=\half$ and the spinor (dimension $\Fr32$) at $\ell=1$. 
The subcomplexes under consideration are 
$r_0^2\ra r_1^1\ra r_2$
and 
$r_0^3\ra r_1^2\ra r_2^1\ra r_3$.
Checking the multiplicities of the relevant representations, (10000) and
(00001), in these, we obtain the sequences
$0\ra1\ra0$ and $0\ra1\ra0\ra0$. The components of the cohomology
in these representations and dimensions clearly contain the physical
fields. This can be understood in a traditional framework as removing
degrees of freedom in a superfield gauge transformation (removing
the image from the left) and imposing the vanishing of the field
strength $F_{\a\b}$ (removing the complement of the kernel from the
right).
Analogous considerations tell us that the second cohomology contains
a spinor of dimension $\Fr52$ and a vector of dimension $3$. These are
interpreted as belonging to a current supermultiplet, \ie, fields
entering the right hand sides of the equations of motion.
This goes well together with the observation that modifications of
the theory are introduced by deforming the constraint $F_{\a\b}=0$
\cite{cede-NilssonSYM,cede-GatesVashakidze,cede-CederwallNilssonTsimpisI,cede-CederwallNilssonTsimpisII}.
The relevance of the cohomology is explained by the facts that deformations
introduced by relaxing $F_{\a\b}=0$ have to fulfill the Bianchi identity
(removing the complement of the kernel from the right),
and that relevant deformations are counted modulo field redefinitions
(removing the image from the left).
See also the following section for a fuller discussion.

A complete calculation of the cohomology requires that one considers
all irreducible representations occurring at arbitrary levels.
This quickly becomes untractable to do by hand. 
The method for calculating cohomologies
is by using the program LiE \cite{cede-LiE}. 
The method will be presented
in detail in a forthcoming publication \cite{cede-CederwallNilssonTsimpisCOHO}.
The complete cohomology consists of
\bea
\H^0=(00000)_0&\qquad&\hbox{(gauge transformations)}\\
\H^1=(10000)_1\oplus(00001)_{3/2}&\qquad&\hbox{(fields)}\\
\H^2=(00010)_{5/2}\oplus(10000)_{3}&\qquad&\hbox{(deformations)}\\
\H^3=(00000)_4&\qquad&\hbox{(?)}
\eea
where the subscript indicates dimension.

Similar cohomologies may be calculated for the $D=11$ supergravity
\cite{cede-CederwallNilssonTsimpisCOHO},
and they confirm in a nice way the conclusions presented earlier in 
this lecture. An interesting observation is that one can choose either
to consider the vielbein or the 3-form, and in either case are all
the fields and deformations of the supergravity contained. It looks as
though a superspace 3-form potential automatically contains gravitational
degrees of freedom, although it is difficult to envisage how the
dynamics should be formulated without reference to geometry.

\subsection{II.4 $F^4$ terms}

I would like to sketch how the superspace methods already described
are used to derive $\a'$-corrections to $D=10$ super-Yang--Mills.
The method for $D=11$ supergravity is in principle analogous, but
much more complicated.
So far, the corrections allowed by supersymmetry have been determined
up to order $\a'^2$ \cite{cede-CederwallNilssonTsimpisII}, and although
the level of technical complexity is high, it seems reasonable to
continue one or two levels.

We need to specify what $J_{abcde}$ is in terms of the fundamental
superfields $F$ and $\l$.
We first observe that there are no corrections at order $\a'$. For
dimensional reasons, $F_{\a\b}$ has to be proportional to $\l^2$, which
does not contain the representation (00020).
Then, starting at order $\a'^2$, there are two types of possible terms, modulo
the lowest order field equations ($A,B,\ldots$ are adjoint 
gauge group indices, not to be confused with $A=(a,\a)$ used earlier): 
\bea
J^A_{abcde}&=&-\half\a'^2 M^A{}_{BCD}(\l^B\G^f\G_{abcde}\G^g\l^C)F^D_{fg}
\nonumber\\
&&+\fr6\a'^2 N^A{}_{BC}\left(D_{[a}\l^B\G_{bcd}D_{e]}\l^C-\hbox{ dual}\,\right)
\punkt\label{cede-JinFields}
\eea
These satisfy the (00030) constraint at linear order, which is
easily seen by acting with a spinor derivative and perform 
tensor multiplication of the representations of the fields.
Here, $M$ and $N$ are some invariant tensors carrying adjoint
indices of the gauge group.

Not all deformations in (00020) are relevant, as explained in the
previous section. Those that are in the image of $\D_1$ correspond
to field redefinitions of $A_\a$ and are trivial. A careful examination
of field redefinitions shows that only the first term in eq. 
(\ref{cede-JinFields}) is relevant, and the other can be discarded. In
addition, $M_{ABCD}$ can be taken to be completely symmetric in
adjoint indices. 

A lengthy calculation gives the deformed equations of motion at order
$\a'^2$ by acting with spinor derivatives on $J_{abcde}$, and inserting
in eq. (\ref{cede-DeformedEOM}). These may subsequently be integrated to a
component action,
which reads
\bea
\L&=&-\fr4G^{Aij}G^A_{ij}+\half\x^A\Dslash\x^A\\
&&-6\a'^2M_{ABCD}\Bigl[\tr G^AG^BG^CG^D-\fr4(\tr G^AG^B)(\tr G^CG^D)\\
&&\qquad
-2G^{A\,i}{}_kG^{B\,jk}(\x^C\G_iD_j\x^D)			
+\fr2G^{A\,il}D_lG^{B\,jk}(\x^C\G_{ijk}\x^D)	\\	
&&\qquad
+\fr{180}(\x^A\G^{ijk}\x^B)(D_l\x^C\G_{ijk}D^l\x^D)		
+\Fr3{10}(\x^A\G^{ijk}\x^B)(D_i\x^C\G_jD_k\x^D)\\		
&&\qquad
+\Fr7{60}f^D{}_{EF}G^{A\,ij}(\x^B\G_{ijk}\x^C)(\x^E\G^k\x^F)\\		
&&\qquad
-\fr{360}f^D{}_{EF}G^{A\,ij}(\x^B\G^{klm}\x^C)(\x^E\G_{ijklm}\x^F)\Bigr]
	+O(\a'^3)\punkt
\eea
The spinor $\l$ has been replaced by $\x$ and $F$ by $G$, since there
is a field redefinition involved in reaching this final form.
It agrees with previous work \cite{cede-BergshoeffFFOUR} on previously
known terms (up to quadratic in fermions).

With only a minor further restriction on $M$, the action has a second
non-linearly realised supersymmetry when the gauge group has a
U(1) factor, as is the case when one considers field theory on
multiple branes. The ``symmetrised trace prescription'' of
Tseytlin \cite{cede-Tseytlin} is consistent with our results, but supersymmetry
does not completely specify it, even at the $F^4$ level.
It will of course be interesting to continue the analysis to higher
orders. The (00030) Bianchi identity will necessarily lead to corrections
at order $\a'^4$ and higher, and a complete action will be non-polynomial.
It it is not clear whether any closed, Born--Infeld-like form exists.
It is even not known if new ``invariants'' arise that start at higher orders,
or if everything follows uniquely once the $\a'^2$ correction is 
determined. 

\subsection{II.5 Branes? Conclusions}

The properties of the maximally supersymmetric field theories
we have considered have been turned into a tool for studying
restrictions imposed by supersymmetry on self-interactions.
Much more is to be done, both for super-Yang--Mills and supergravity,
but it will be necessary to use computer programs, \eg\ LiE
\cite{cede-LiE} and the Mathematica package GAMMA \cite{cede-GAMMA},
to a higher degree.

A question which so-far remains unaddressed is what happens to branes
moving in backgrounds with $\a'$-corrections from string/M-theory.
To investigate this one will need more informations about 
$\a'$-corrected supergravity. 
Will the actions still be formally the same, and the dynamics only
change through the coupling to background fields? I would tend to
answer in the positive, although nothing is known. One difficulty
immediately presents itself, namely that the tensor field strengths
will take non-zero values even for the components of negative dimension
\cite{cede-CGNN}.
Since $\k$-symmetry relies on cancellations of contributions from
the kinetic and WZ terms, the resulting variations would have no
contribution from the torsion to cancel against. One possibility is
that also the condition that $\k$ is purely spinorial is modified.
One preliminary investigation would consist of checking $\k$-symmetry
for supersymmetric Wilson loops \cite{cede-HolographyInSS} in a
deformed super-Yang--Mills background.

\begin{theacknowledgments}
The author would like to thank the organisers of the XXXVII Karpacz
Winter School for two very pleasant and inspiring weeks.
\end{theacknowledgments}

                                                          %\hbox to\hsize{\hfill\tiny S.D.G.}           

\begin{thebibliography}{8}                                                                             
\expandafter\ifx\csname natexlab\endcsname\relax\def\natexlab#1{#1}\fi                                 
\providecommand{\enquote}[1]{``#1''}                                                                   
\expandafter\ifx\csname url\endcsname\relax                                                            
  \def\url#1{\texttt{#1}}\fi                                                                           
\expandafter\ifx\csname urlprefix\endcsname\relax\def\urlprefix{URL }\fi                               
\bibitem{cede-Stelle}{K.S. Stelle,
{\xit ``BPS branes in supergravity''},
\hepth{9803116}.}

\bibitem{cede-Duff}{M.J. Duff,
{\xit ``TASI lectures on branes, black holes and anti-de Sitter space''},
\hepth{9912164}.}

\bibitem{cede-Achucarro}{A. Achucarro, J.M. Evans, P.K. Townsend and D.L. Wiltshire,
{\xit ``Super $p$-branes''},
\PLB{198}{1987}{441}.}

\bibitem{cede-BSTMii}{E. Bergshoeff, E. Sezgin and P.K. Townsend, 
{\xit ``Supermembranes and eleven-dimensional supergravity''},
\PLB{189}{1987}{75};
{\xit ``Properties of the eleven-dimensional supermembrane theory''},
\AP{185}{1988}{330}.}

\bibitem{cede-DuffLu}{M.J. Duff and  J.X. Lu,
{\xit ``Type II $p$-branes: the brane scan revisited''},
\NPB{390}{1993}{276} [\hepth{9207060}].}

\bibitem{cede-HoweSezginSuperBranes}{P.S. Howe and E. Sezgin,
{\xit ``Superbranes''},
\PLB{390}{1997}{133} [\hepth{9607227}].} 

\bibitem{cede-Dp}{M. Cederwall, A. von Gussich, B.E.W. Nilsson and A. Westerberg,
{\xit ``The Dirichlet super-three-brane in ten-dimensional type IIB 
supergravity''}
\NPB{490}{1997}{163} [\hepth{9610148}];\nlni
M. Aganagi\'c, C. Popescu, J.H. Schwarz,
{\xit ``D-brane actions with local kappa symmetry''},
\PLB{393}{1997}{311} [\hepth{9610249}];\nlni
M. Cederwall, A. von Gussich, B.E.W. Nilsson, P. Sundell
 and A. Westerberg,
{\xit ``The Dirichlet super-p-branes in ten-dimensional type IIA and IIB 
supergravity''},
\NPB{490}{1997}{179} [\hepth{9611159}];\nlni
E. Bergshoeff and P.K. Townsend, 
{\xit ``Super D-branes''},
\NPB{490}{1997}{145} [\hepth{9611173}].}

\bibitem{cede-Polchinski}{J. Polchinski,
{\xit ``Dirichlet branes and Ramond-Ramond charges''},
\PRL{75}{1995}{4724} [\hepth{9510017}].} 

\bibitem{cede-ElevenSG}{E. Cremmer, B. Julia and J. Sherk, 
{\xit ``Supergravity theory in eleven-dimensions''},
\PLB{76}{1978}{409}.}

\bibitem{cede-ElevenSSSG}{L. Brink and P. Howe, 
{\xit ``Eleven-dimensional supergravity on the mass-shell in superspace''},
\PLB{91}{1980}{384};
E. Cremmer and S. Ferrara,
{\xit ``Formulation of eleven-dimensional supergravity in superspace''},
\PLB{91}{1980}{61}.}

\bibitem{cede-HoweWeyl}{P.~Howe,
{\xit ``Weyl superspace''},
\PLB{415}{1997}{149} [\hepth{9707184}].}

\bibitem{cede-DuffKhuriLu}{M.J. Duff, R.R. Khuri and J.X. Lu,
{\xit ``String solitons''},
\PR{259}{213}{1995} [\hepth{9412184}].}

\bibitem{cede-ACGNR}{T. Adawi, M. Cederwall, U. Gran, B.E.W. Nilsson and
	B. Razaznejad,
{\xit ``Goldstone tensor modes''},
\JHEP{99}{02}{1999}{001} [\hepth{9811145}].} 

\bibitem{cede-CGNN}{M. Cederwall, U. Gran, M. Nielsen and B.E.W. Nilsson, 
{\xit ``Manifestly supersymmetric M-theory''}, 
\JHEP{00}{10}{2000}{041} [\hepth{0007035}];
{\xit ``Generalised 11-dimensional supergravity''}, \hepth{0010042}.}

\bibitem{cede-GatesNishino}{S.J. Gates, Jr. and H. Nishino, 
{\xit ``Deliberations on 11D superspace for the M-theory effective action''},
\hepth{0001037}.}

\bibitem{cede-Leigh}{R.G. Leigh, 
	{\xit ``Dirac--Born--Infeld action from Dirichlet sigma
	model''}, Mod.~Phys.~Lett.~{\xbf A4} ({\xold1989}) {\xold2767};\nlni
	C.G. Callan, C. Lovelace, C.R. Nappi and S.A. Yost,
	{\xit ``String loop corrections to beta functions''},
	\nl Nucl.~Phys. {\xbf B288} ({\xold1987}) {\xold525}.}

\bibitem{cede-Douglas}{M.~Douglas, {\xit ``Branes within branes''},
	\hepth{9512077};\nlni
M.B.~Green, C.M.~Hull and P.K.~Townsend, \nl{\xit ``D-brane 
	Wess--Zumino actions, T-duality and the cosmological constant''},
	Phys. Lett. {\xbf B382} ({\xold1996}) {\xold65} 
	[\hepth{9604119}].}

\bibitem{cede-FiniteFSols}{J.M.~Izquierdo, N.D.~Lambert, G.~Papadopoulos and
P.K.~Townsend, ``Dyonic membranes'', Nucl.\ Phys.\ {\bf B460} (1996)
560 [hep-th/9508177];\nlni
J.G.~Russo and A.A.~Tseytlin, ``Waves,
boosted branes and BPS states in M-theory'', Nucl.\ Phys.\ {\bf B490}
(1997) 121 [hep-th/9611047];\nlni
M.~Cederwall, U.~Gran, M.~Holm and B.E.W.~Nilsson,
``Finite tensor deformations of supergravity solitons'', JHEP {\bf
9902} (1999) 003 [hep-th/9812144];\\
M.~Cederwall, U.~Gran, M.~Nielsen, and B.E.W. Nilsson, ``$(p,q)$
5-branes in
non-zero B-field'', JHEP {\bf0001} (2000) 037 [hep-th/9912106].}

\bibitem{cede-CarrGatesOerterIIA}{J.L. Carr, S.J. Gates, Jr. and R.N. Oerter,
{\xit ``$D=10$, $N=2A$ supergravity in superspace''},
\PLB{189}{1987}{68}.} 

\bibitem{cede-HoweWestIIB}{P.S. Howe and P.C. West,
{\xit ``The complete $N=2$, $D=10$ supergravity''},
\NPB{238}{1984}{181}.} 

\bibitem{cede-CT}{M. Cederwall and P.K. Townsend, {\xit ``The manifestly 
	{\xrm SL(2;{\tenbbb Z})}-covariant superstring''},
	\nl\jhep{97}{09}{1997}{003} [\hepth{9709002}]}

\bibitem{cede-CW}{M. Cederwall and A. Westerberg,
	{\xit ``World-volume fields, {\xrm SL(2;{\tenbbb Z})} and
		duality: the type \II B 3-brane''},
		\nl\jhep{98}{02}{1998}{004} [\hepth{9710007}].}

\bibitem{cede-WittenDbranes}{E. Witten, 
{\xit ``Bound states of strings and $p$-branes''},
\NPB{460}{1996}{335} [\hepth{9510135}].}

\bibitem{cede-SchwarzSL}{J.H. Schwarz, 
	{\xit ``An {\xrm SL(2;Z)} multiplet of 
	type IIB superstrings''}, 
	Phys. Lett. {\xbf B360} ({\xold 1995}) {\xold 13}
	\nl[\hepth{9508143}];
        Erratum: ibid. {\xbf B364} ({\xold 1995}) {\xold 252}.} 

\bibitem{cede-CGNNpq}{M.~Cederwall, U.~Gran, M.~Nielsen, and B.~E.~W. Nilsson, 
``$(p,q)$ 5-branes in
non-zero B-field'', JHEP {\bf0001} (2000) 037 [hep-th/9912106].}

\bibitem{cede-SchwarzSixDual}{M. Aganagic, J. Park, 
C. Popescu and J.H. Schwarz,
{\xit ``Dual D-brane actions''},
\NPB{496}{1997}{215} [\hepth{9702133}].}

\bibitem{cede-CNS}{M. Cederwall, B.E.W. Nilsson and P. Sundell, 
	{\xit ``An action for the 5-brane in $D=11$ supergravity''},
	\nl\jhep{04}{1998}{007} [\hepth{9712059}].} 

\bibitem{cede-BBG}{C.P.~Bachas, P. Bain and M.B.~Green, 
{\xit ``Curvature terms in the D-brane actions and their M-theory origin''},
\JHEP{05}{1999}{11} [\hepth{9903210}].}

\bibitem{cede-Wyllard}{N. Wyllard,
{\xit 
``Derivative corrections to D-brane actions with constant background fields''},
\hepth{0008125}.}

%\bibitem{cede-ConvConstr}{S.J.~Gates, K.S.~Stelle and P.C.~West,
%{\xit ``Algebraic origins of superspace constraints in supergravity''},
%\NPB{169}{1980}{347}; 
%S.J. Gates and W. Siegel, 
%{\xit ``Understanding constraints in 
%superspace formulation of supergravity''},
%\NPB{163}{1980}{519}.}

\bibitem{cede-BNLXVI}
{B.E.W.~Nilsson, {\xit ``A supersymmetric approach to branes and 
supergravity''}, \xrm in ``Theory of elementary particles'', Proc. of the
{\xold31}st international symposium Ahrenshoop, September {\xold2}-{\xold6}, 
{\xold1997}, Buckow,
Eds H. Dorn et al. (Wiley-VCH {\xold1998}), 
G\"oteborg-ITP-{\xold98}-{\xold09} [\hepth{0007017}].}

\bibitem{cede-Tseytlin}{A.A.~Tseytlin, 
{\xit ``Born--Infeld action, supersymmetry and string theory''},
\xrm in the Yuri Golfand memorial volume, ed. M. Shifman,
World Scientific (2000) [\hepth{9908105}];
{\xit ``On the non-abelian generalization of 
Born--Infeld action in string theory''},
\NPB{501}{1997}{41} [\hepth{9701125}].}

\bibitem{cede-OtherTalks}{See also the talks by Ivanov, Krivonos and
Bergshoeff at this school.}

\bibitem{cede-NilssonSYM}{B.E.W.~Nilsson, 
\xit ``Off-shell fields for the 10-dimensional supersymmetric 
Yang--Mills theory'', \xrm G\"oteborg-ITP-{\xold81}-{\xold6};
{\xit ``Pure spinors as auxiliary fields in the ten-dimensional 
supersymmetric Yang--Mills theory''},
\CQG3{1986}{{\xrm L}41}.}

\bibitem{cede-CederwallNilssonTsimpisI}
{M. Cederwall, B.E.W. Nilsson and D. Tsimpis,
{\xit ``The structure of maximally supersymmetric gauge theories: 
constraining higher order interactions''}, \hepth{0102009}.}

\bibitem{cede-LiE}{A.M. Cohen, M. van Leeuwen and B. Lisser, 
LiE v. {\xold2}.{\xold2} ({\xold1998}), 
\nlni http://wallis.univ-poitiers.fr/\~{}maavl/LiE/} 

\bibitem{cede-GatesVashakidze}{S.J.~Gates, Jr. and Sh.~Vashakidze
{\xit ``On $D=10$, $N=1$ supersymmetry, superspace geometry and 
superstring effects''},
\NPB{291}{1987}{172}.}

\bibitem{cede-CederwallNilssonTsimpisII}
{M. Cederwall, B.E.W. Nilsson and D. Tsimpis,
{\xit ``$D=10$ super-Yang--Mills at $O(\a'^2)$''}, \hepth{0104236}.}

\bibitem{cede-CederwallNilssonTsimpisCOHO}
{M. Cederwall, B.E.W. Nilsson and D. Tsimpis,
{\xit in preparation}.}

\bibitem{cede-BergshoeffFFOUR}{E.~Bergshoeff, M.~Rakowski and E.~Sezgin,
{\xit ``Higher derivative super-Yang--Mills theories},
\PLB{185}{1987}{371}.}

\bibitem{cede-GAMMA}{U. Gran,
{\xit ``GAMMA: A Mathematica package for performing gamma-matrix 
algebra and Fierz transformations in arbitrary dimensions''},
\hepth{0105086}.}

\bibitem{cede-HolographyInSS}{H. Ooguri, J. Rahmfeld, H. Robins and J. Tannenhauser,
{\xit ``Holography in superspace''},
\JHEP{00}{07}{2000}{045} [\hepth{0007104}].}

\end{thebibliography}
                                                                               \end{document}